# RELEVANCE OF THE SPEED AND DIRECTION OF PULLING IN SIMPLE MODULAR PROTEINS

*Carlos A. Plata[1], Zackary N. Scholl[2], Piotr E. Marszalek[3], and Antonio Prados[1]\**

[1] Física Teórica, Universidad de Sevilla, Apdo. de Correos 1065, Sevilla 41080, Spain.

[2] Department of Physics, University of Alberta, Edmonton, Alberta, Canada.

[3] Department of Mechanical Engineering and Materials Science, Duke University, Durham, North Carolina.

ABSTRACT

A theoretical framework capable of predicting the first unit that unfolds in pulled modular proteins has been recently introduced, for "fast enough" pulling velocities. Within this picture, we investigate the unfolding pathway in a chain of identical units and predict that the module closest to the pulled end opens first. Steered molecular dynamics of a simple construct, specifically a chain composed of two coiled-coil motives, shows that this is indeed the case. Notwithstanding, the unfolding behavior strongly depends on the terminus (C or N) from which this homopolyprotein is pulled. Therefore, anisotropic features are revealed and seem to play an important role for the observed unfolding pathway.

ABSTRACT GRAPHIC:

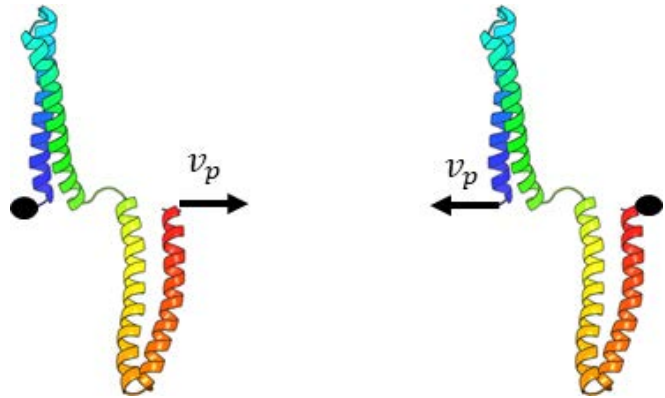

INTRODUCTION

Protein folding and unfolding plays a central role in many biological processes. The emergence of single molecule manipulation techniques, mainly Laser Optical Tweezers and Atomic Force Microscopy (AFM), has made it possible to improve our understanding of elastic properties of protein systems and other biomolecules[1–4]. Specifically, a large effort has been devoted to the analysis of the so-called force-extension curves, obtained when pulling the biomolecule by either controlling its end-to-end distance or the applied force[5–8].

Statistical mechanics has been shown to be a powerful tool to investigate the elastic behavior of proteins and other biomolecules. In particular, the Jarzynski identity[9,10] and the Crooks fluctuation theorem[11] have been used to reconstruct relatively simple free energy landscapes from non-equilibrium experiments in systems with two well-defined minima separated by a barrier[12,13]. Also, recent developments such as the Minh-Adib's bidirectional estimator[14] have also made it possible to obtain the free energy surface in more complex cases, with many local barriers and minima[15,16]. Some of the most recent developments in this enticing field are discussed in a recent review by Seifert[17].

It is many small proteins that unfold in a two-state manner, being either folded or unfolded with no significant intermediate states. In contrast, larger proteins often comprise several stable subunits or "unfoldons" (using the terminology introduced by Bertz and Rief[18]), which unfold independently of the remainder of them. Thus, the unfolding of these larger proteins is composed of several steps; unraveling one of the unfoldons in each of those steps. Interestingly, this unfolding pathway, i.e. the order and the way in which these subunits are opened, depends on both the velocity and the direction of pulling[19–23].

AFM experiments allowed Bertz and Rief to characterize four of these unfoldons in the Maltose-Binding Protein (MBP). They labeled these unfoldons as M1, M2, M3 and M4, with M1 being the closest to the C-terminus and M3 the closest to the N-terminus. Pulling MBP at a typical speed of $10^{-9}$ nm/ps, Bertz and Rief found a well-defined unfolding pathway: the weakest unfoldon (M1), i.e. that characterized by the lowest opening force, unraveled first. Thereafter, the remainder of the unfoldons opened sequentially, from weakest to strongest (M4).

Later, Guardiani et al. had a more detailed look into the unfolding of MBP. They showed[22], by means of a combination of Gō model simulations and steered molecular dynamics, that the unfolding pathway is more complex and seems to depend both on the velocity and direction of pulling. C-pulling simulations of the Gō model always showed a pathway compatible with Bertz and Rief's experiment. However, N-pulling simulations of the Gō model displayed a different behavior: for small velocities, again Bertz and Rief's pathway was found, but for high enough pulling speed it was M3, the closest to the N-terminus, that opened first. Steered molecular dynamics simulations of MBP at a pulling speed of $5\times10^{-3}$ nm/ps gave results that were consistent with those from the Gō model at high velocities, showing the different pathways (M1 vs. M3 opening first) depending on the pulled terminus (C vs. N).

A deeper understanding of the above results can be obtained within a quite general theoretical framework. The typical approach is to consider that each of the biomolecule's structural units can be modeled by a particle moving in an effective potential or free energy landscape[6–8,24–26]. In a certain range of forces, the free energy has two minima corresponding to different system lengths $l(F)$ and $(U)$ ($l(F) < l(U)$), see left panel in Figure 1. Naturally, $F$ and $U$ stand for "folded" and "unfolded", respectively. Thus, as shown in the right panel of Figure 1, a multistability region appears in the equilibrium branches of the force-extension curve[7]. Within this range, each unit unfolds by jumping from the minimum corresponding to $l(F)$ to the minimum corresponding to $l(U)$ at a value of the force that depends on the pulling velocity[8,27].

Let us focus on the first unfolding event for pulling processes in which the end-to-end length of the biomolecule is the controlled quantity. If pulling is very slow and quasistatic, the first unfolding event occurs at the length value for which the free energy minima over the branch with all the units folded and over the branch with only one unit unfolded are equally deep[7]. Therefore, the jump between branches occurs by thermal activation over the free energy barrier separating them. Then, the completely folded branch is only partially swept, as marked by the dashed vertical (red) lines in the right panel of Figure 1. In contrast, there is a range of fast pulling velocities that do not give the system enough time to be thermally activated over the barrier, but are slow enough to allow it to sweep completely the part of the branches that corresponds to metastable equilibrium states. In this case, the jump between branches comes about at the limit of metastability, only when the folded minimum disappears[8], as marked by the solid vertical (blue) lines in the right panel of Figure 1. This range of velocities has been called "adiabatic"[8] or said to lead to the "maximum hysteresis path"[25], and it is the velocity range in which we are interested in this work.

The above presented theoretical framework has been employed to investigate the unfolding pathway of proteins comprising independent unfoldons, like MBP. Therein, the main theoretical result can be summarized as follows: for slow pulling, it is the weakest unit that opens first but for fast enough pulling, it is the unit closest to the pulled end that opens first[23]. This explains Guardiani et al.'s numerical results for the pulling of MBP: in C-pulling experiments, the weakest and the closest to the pulled end unit coincide (M1), and thus the unfolding pathway is always identical to that previously observed by Bertz and Rief, whereas in N-pulling experiments, the closest to the pulled end unit is M3, and therefore the first unit that unfolds changes from M1 to M3 as the pulling speed is increased.

Clearly, not only can the above picture be applied to proteins comprising several unfoldons but also to modular proteins, in which the structural units are naturally associated with the modules. In this respect, the analysis of the unfolding pathway has been scarcely studied, probably because the order in which the modules are opened cannot be directly elucidated from current experiments. Therefore, there are still a lot of unanswered questions, which we attempt to clarify, at least partially, in this work.

In order to keep things as simple as possible, we consider here a homopolyprotein, i.e. a modular protein in which all the units are identical. In order to predict their unfolding pathway, we employ a recently developed theory for biomolecules comprising several units[23], but restricting it to the specific case of homopolyproteins. Therefore, we make use of a Langevin description within the so-called macroscopic approximation[28]. Our main prediction is that the unit closest to the pulled end always opens first, as long as the velocity is "fast enough", in the sense discussed above. Moreover, we test this prediction by carrying out steered molecular dynamic (SMD) simulations in a particularly simple system composed of coiled coils. This kind of structures is common in nature, which makes it extremely useful as a model system[29–31].

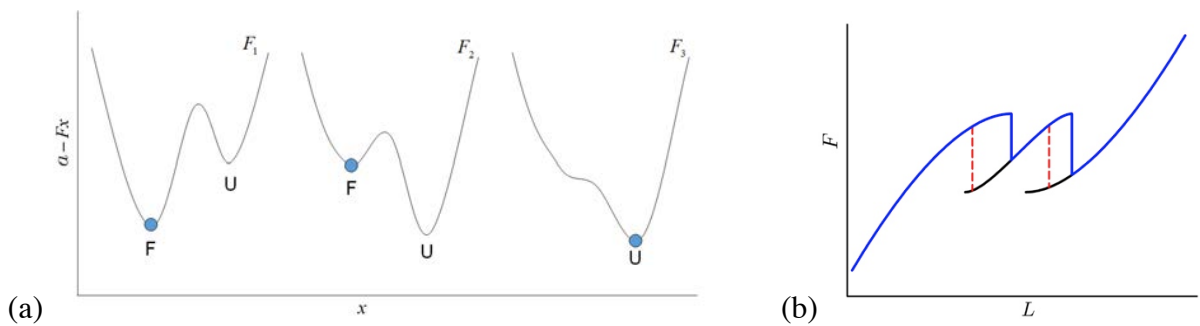

**Figure 1.** (a) Schematic free energy landscape of a single repeat for three different forces $F_1 < F_2 < F_3$. The system starts in the $F$ state, which is the absolute minimum for $F_1$. If thermal noise is neglected, the repeat remains in the folded state for $F_2$ even when the unfolded state is more stable. For $F_3$, the $F$ state disappears and the repeat finally unfolds. (b) Qualitative picture of the stability branches in a modular system with two units. The blue line stands for the unfolding pathway followed in the limit of the "maximum hysteresis path", when the pulling speed is high enough to make the system sweep the whole branches. In this limit, the jumps between consecutive branches take place by the mechanism shown in the left panel because the system does not have enough time to jump over the barrier separating the folded and unfolded states, i.e. the "fast enough" pulling speed effectively suppresses thermal. Conversely, in the quasistatic limit the transition from the folded to the unfolded basin occurs at the lengths (dashed red lines) at which the branch with one more unfolded unit becomes more stable, that is, when its free energy becomes smaller; therein the system always have time to find its way through the barrier.

## MODEL AND METHODS

### Simple Model

We consider a homopolyprotein composed of $N$ non-interacting identical repeats. When the molecule is submitted to a pulling force along a defined axis, the simplest description is one-dimensional. As depicted in Figure 2, we define the coordinates $q_i$, $(i = 0,1,...,N)$, in such a way that the $i$-th repeat extends from $q_{i-1}$ to $q_i$. Thus, the extension of the $i$-th repeat can be found as a simple subtraction $x_i = q_i - q_{i-1}$.

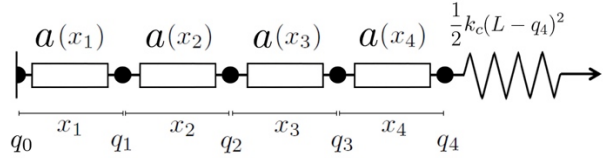

**Figure 2.** Sketch of a homopolyprotein of 4 repeats. Each repeat contributes to the free energy with a function that only depends on its extension. There is an extra elastic contribution from the pulling device that tries to keep the total length of the molecule equal to $L$.

The dynamics of the system is governed by the set of Langevin equations

$$\gamma \dot{q}_i = -\frac{\partial}{\partial q_i} A(q_0, q_1, ..., q_N) + \zeta_i, \quad i > 0; \quad q_0 = 0. \quad (1)$$

Therein, $\zeta_i$ represent Gaussian white noise forces verifying the fluctuation-dissipation theorem, i.e.

$$\langle \zeta_i(t) \rangle = 0, \quad \langle \zeta_i(t)\zeta_j(t') \rangle = 2\gamma k_B T \delta_{i,j} \delta(t - t'). \quad (2)$$

Above, $k_B$ is the Boltzmann constant, and $\gamma$ and $T$ are the friction coefficient and the temperature of the fluid in which the molecule is immersed, respectively. As usual, the symbol $\langle \cdot \rangle$ in eq (2) stands for the average over the noise. We assume the global free energy of the system to be

$$A(q_0, q_1, \ldots, q_N) = \sum_{i=1}^{N} a(q_i - q_{i-1}) + \frac{1}{2}k_c(L - q_N)^2. \quad (3)$$

The previous equation assigns to each repeat a contribution $a$, which only depends on the extension, and an elastic term that tries to keep the total length of the molecule $q_N$ equal to the desired program $L(t)$. In this work, we focus on length-controlled experiments at constant rate $v_p$, therefore $dL/dt = v_p$.

Note that when $k_c \to \infty$ the control over the length is perfect and $q_N \to L$ for all times in such a way that $k_c(L - q_N) \to F$, becoming a Lagrange multiplier that assures that the constraint over the total length holds. In order to keep the model as simple as possible, we consider (i) perfect control over the length and (ii) the so-called macroscopic approximation[28], that is, we neglect the noise terms in the Langevin approach. The validity and physical meaning of this approximation is discussed later. Note that the system evolution still depends on the temperature, because the single-repeat free energy $a$ contains the temperature.

Taking into account the above discussed approximations, we reach the following evolution equations for the extensions

$$\gamma \dot{x}_1 = -a'(x_1) + a'(x_2), \quad (4a)$$

$$\gamma \dot{x}_i = -2a'(x_i) + a'(x_{i+1}) + a'(x_{i-1}), \quad 1 < i < N, \quad (4b)$$

$$\gamma \dot{x}_N = -2a'(x_N) + a'(x_{N-1}) + F, \quad (4c)$$

$$F = \gamma v_p + a'(x_N). \quad (4d)$$

The reader familiarized with AFM experiments may note that our model does not completely match the usual AFM setup, in which the elastic force stemming from the bending of the cantilever is located on the opposite end (see Figure 1 of Marszalek and Dufrêne's work[2]), but that in steered molecular dynamics simulations. Nevertheless, in the limit of perfect length control considered here, the evolution equations are independent of the location of the elastic reaction.

Up to this moment, we have made no assumptions about the particular shape of the contribution of a single repeat to the free energy $a$. In a simplifying view, each repeat has two states, folded ($F$) and unfolded ($U$). Therefore, they can be represented by a potential $a(x)$ if the function $a - Fx$ has a double-well shape for some interval of forces, see left panel of Figure 1. In this interval, the chain presents a metastable behaviour[7]. For low values of the force the $F$ state has a lower energy than the $U$ state, making the first one more stable. Increasing the force, we start making the $U$ state more and more stable until the $F$ state ceases to exist.

The dynamical eqs (4a) can be solved by means of a perturbative expansion in the pulling speed $v_p$. The perturbative solution, up to a linear order, starting from an initial condition where all the repeats are folded, is[23]

$$x_i = l_s + \gamma v_p \frac{3i(i-1) - (N^2 - 1)}{6Na''(l_s)}, \quad (5)$$

where $l_s = L/N$ is the specific length per repeat. According to eq (5), the closer a repeat is to the pulled end (greater $i$), the faster its length is increased. The idea is that the first unit that unfolds is the first that reaches the stability threshold $\ell$, which marks the end of the metastability region. Precisely, $a''(\ell) = 0$ and therefore the approximate solution given by eq (5) would cease to be valid as the stability threshold is approached. However, we expect this solution to be useful along the initial evolution of the system and, specifically, to predict which repeat is the first to unfold. In a homopolyprotein, the answer seems to be simple within this approach: the most elongated repeat is the pulled one (the closest to the moving end of the pulling device) and thus this will be the first to unfold.

The main assumption in our theory is the macroscopic approximation, i.e. neglecting the thermal noise terms in the Langevin equations. This means that the only way to surpass from $F$ to $U$ state is to reach the limit of stability, at which the folded minimum disappears, as qualitatively shown in left panel of Figure 1. Of course, the accuracy of this approximation improves as the temperature is lowered, but this is not the way in which it is expected to arise in pulling experiments. As discussed in the introduction, this approximation is expected to be valid as long as the pulling velocity is slow enough to allow the system to sweep the metastable region of the equilibrium branches but not so slow to allow the system jump over the barrier separating the folded and unfolded states by thermal activation. Therefore, the system completely sweeps the metastable region of the equilibrium branches and the "adiabatic or deterministic limit", using the terminology of Bonilla et al.[8], or the "maximum hysteresis path", using the terminology of Benichou et al.[24,25], is attained. In other words, we are considering a range of "fast enough" or "adiabatic"[8] pulling speeds leading to the "maximum hysteresis path"[25].

In summary, our theory, which we have briefly developed above for a homopolyprotein and discussed in more detail and for a more general biomolecule in a previous work[23], predicts that the first repeat to unfold in a homopolyprotein would be the pulled one. The main assumption is that thermal activation over the free energy barrier separating the folded and unfolded basins can be neglected, as a consequence of the considered range of pulling velocities. The unimportance of thermal activation processes entails that the unfolding pathway is mainly deterministic, as described above.

**Candidate to test**

Our theoretical approach is clearly a drastic simplification of reality. In fact, our theory is deterministic, in the sense that the unfolding pathway is a definite one, the randomness coming from thermal fluctuations being effectively "suppressed" by the fast enough pulling velocity. In reality, the unfolding pathway does have some degree of stochasticity, stemming from the interactions between the molecule under study and the fluid where it is immersed, which are encoded in the Gaussian white noises of the Langevin description.

Following the discussion above, one expects our theoretical approach to hold for some molecules within a specific range of pulling velocities. One of the obvious requirements our protein candidate must meet is a negligible interaction between repeats, since we have assumed no nearest-neighbor interaction terms in the global free energy. Regarding the range of velocities, we need to be in the regime of the maximum hysteresis path, in which the unfolding of a repeat comes about because its extension has reached its limit of stability.

We have designed a simple homopolyprotein, which we employ below to test whether it fits our theoretical description. We have extracted the structure of an antiparallel coiled-coil motif (CC) from the archeal box C/D sRNP core protein (Protein Data Bank entry 1NT2), which comprises 67 residues and whose N-terminus and C-terminus are, respectively, arginine and isoleucine[32]. This structure has been proven to be useful as a mechanical folding probe[33]. We use this CC as the building blocks of the molecule: our system is simply a concatenation of two CC motives connected by a linker, which is composed of two consecutive pairs of alternated residues of glycine and serine. We expect this linker not to introduce any significant interaction between the two domains. The initial conformation of the constructed model structure and orientation of the two CC repeats is shown in Figure 3. The end-to-end vector points from the N-terminus to the C-terminus, aligned with the $x$-axis, whereas both axial directions of the two CC structures are located as parallel as possible to the $z$-axis.

According to the theoretical framework we have developed, we expect that if we pull from one end of the designed molecule, the first repeat to unfold will be precisely the closest to the moving end. To test this theory, we perform SMD simulations to analyze the degree of agreement of the obtained numerical results with the above developed theory.

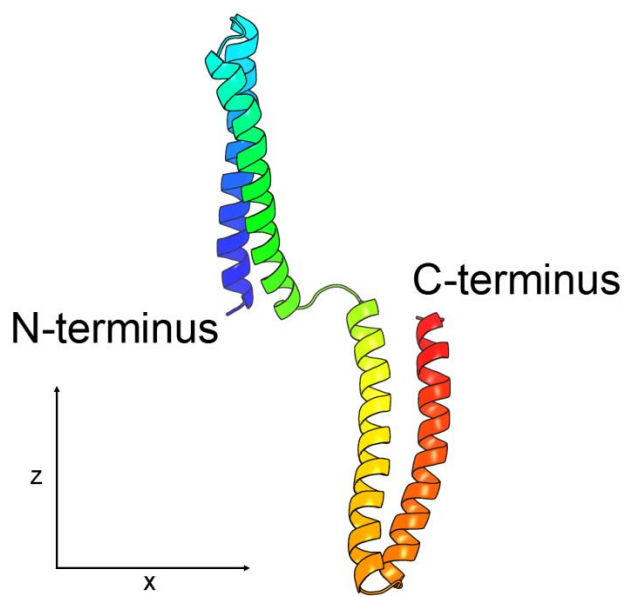

**Figure 3.** Initial conformation of the homopolyprotein composed of two CCs in the SMD simulations. The pulling direction is aligned with the $x$-axis, whereas the axial directions of the CCs are aligned with the $z$-axis.

**All-atom Molecular Dynamics Simulation**

In order to minimize any technical difficulty stemming from experimental setup, we restrain the analysis in this work to all-atom molecular dynamics simulations. Such simulations start from the initial conformation shown in Figure 3. First, we add hydrogen atoms using VMD Automatic PSF Builder[34]. Then, we create a water box of size 300 Å × 70 × Å × 120 Å, in the $x$-, $y$-, and $z$-axes, respectively. Note that this box is long enough in the direction of pulling, i.e. the $x$-axis, to contain the unfolded protein. Also, NaCl is introduced in the system replacing water molecules until the concentration reaches 150 mM/L and the charge is neutralized. Finally, simulations are performed using NAMD2 2.10[35], with two different parts: (i) the "equilibration stage" at 310 K and (ii) the "pulling stage" with velocity $v_0 = 1.4 \cdot 10^{-2}$ nm/ps and a force constant 4860 pN/nm. We have also considered faster pulling velocities, namely $2v_0$ and $5v_0$, as detailed below. The molecule's behavior for these faster pulling velocities has been investigated in order to elucidate whether or not the unfolding pathway becomes more deterministic as the pulling speed is increased.

These typical pulling speeds in steered molecular dynamics simulations are higher by several orders of magnitude than the experimental ones, but they are necessary to investigate this kind of system with the available computer power. In addition, this high velocity range is especially relevant for our present purposes, since we are interested in exploring the maximum hysteresis path limit. Note that the considered value for the stiffness of the elastic reaction is also two orders of magnitude higher than the typical ones in AFM experiments, and thus closer to the perfect length control situation assumed in our theory.

A notable number of pulling trajectories $N_T$ are needed in order to obtain a meaningful statistical analysis of the unfolding pathway. The final configuration of the molecule in the equilibration stage is taken as the initial condition for the pulling stage. In order to generate different initial conditions for pulling, we have considered one "long" trajectory in the equilibration stage and collected the molecule configurations at several different times $t_k$, with $t_{k+1} - t_k \geq 0.1$ ns, as the initial conditions for the different pulling trajectories $k = 1, \ldots, N_T$. We have checked the "independence" of the trajectories obtained from these initial conditions, in the sense that the unfolding pathway from two consecutive initial conditions, corresponding to $k$ and $k + 1$, are not correlated. Details are given in the supporting information that accompanies this paper.

The duration of the pulling stage $\Delta t_p$ is chosen to allow the molecule to unravel. The size of a single CC motif in its axial direction is around 5 nm (4.82 nm between the two Cαs most separated in the axial direction), thus a motif can be considered as completely unfolded when its end-to-end distance (measured between their Cαs in the terminal residues ARG and ILE) exceeds 10 nm. For the "base" velocity $v_0 = 1.4 \cdot 10^{-2}$ nm/ps, we have chosen $\Delta t_p = 1.6$ ns, so that the total length increment is $v_0 \Delta t_p = 22.4$ nm. For the faster pulling velocities, $2v_0$ and $5v_0$, we have decreased $\Delta t_p$ accordingly.

RESULTS AND DISCUSSIONS

In SMD simulations, the length of each repeat can be measured as a function of time and thus we may introduce the basic (and the simplest) classification of trajectories by labeling them as "Good" (G) if the CC motif closest to the pulled end unfolds sooner than the furthest, and "Bad" (B) otherwise. Clearly, it is G-trajectories that agree with the prediction of our theory for identical units. The above basic classification of SMD trajectories as G or B can be refined, so as to have a more accurate description of the unfolding pathway. In our theoretical framework, the unfolding is sequential: when the first unit unfolds, the second one has not reached its limit of stability and thus remains in the folded state. However, in the simulations, this perfectly sequential unfolding is not always found: when the first unit unfolds, the second one can be partially unfolded.

Following the discussion in the previous paragraph and for the sake of accuracy, we define a criterion for distinguishing between different subtypes of trajectories. Therefore, we incorporate a quantitative measurement of the degree of unfolding for each repeat; firstly in a geometric way and secondly taking into account the fraction of native contacts[36]. As already said above, the size of the CC motif in its axial direction is approximately 5 nm and thus we consider a motif to be completely unfolded when its end-to-end distance is greater than 10 nm. Also, we define when a motif is partially unfolded in our simulations by introducing an "unfolding threshold", i.e. a length below which we consider the unit to be still folded. Specifically, we take this unfolding threshold to be 7 nm, which corresponds to an opening angle of 90° in a rigid rods picture, as depicted in Figure 4.

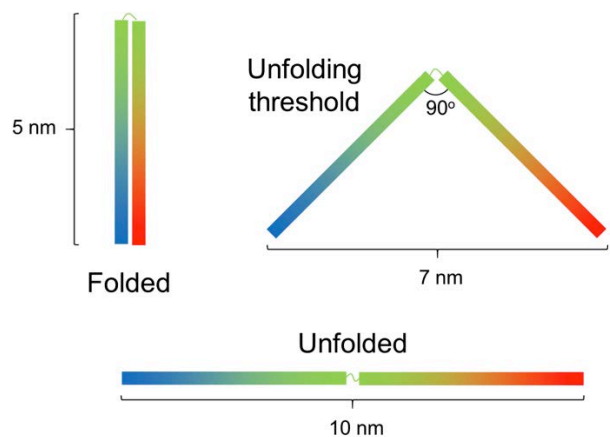

**Figure 4.** Simple unfolding criterion for the CC under study based on a rigid-rod picture. Depending on the value of its end-to-end distance, the molecule is regarded as (i) folded if it is smaller than 7 nm, (ii) partially unfolded when it is between 7 and 10 nm, and (iii) completely unfolded if it is greater than 10 nm.

The above "geometric" choice for the unfolding threshold has some degree of arbitrariness, especially in relation to the length (or angle between rigid-rods) chosen for the unfolding threshold. In order to give a physical basis for this choice, we look into the fraction of native contacts[36] as a function of the total length for a single CC motif in Figure 5. For this purpose, we have performed a SMD simulation in which a single CC motif, which is initially folded, is pulled at a speed of $3.75 \cdot 10^{-3}$ nm/ps from its C-terminus. It can be observed how the number of native contacts decreases along the trajectory, with a well-defined plateau arising between (approximately) 7 and 10 nm. Clearly, the borders of this plateau demarcate the region where the main bonds that keep the double-stranded CC folded are broken. Therefore, the above-defined geometric thresholds for considering the molecule folded/partially unfolded/completely unfolded agree with the corresponding limits in the fraction-of-native-contacts picture.

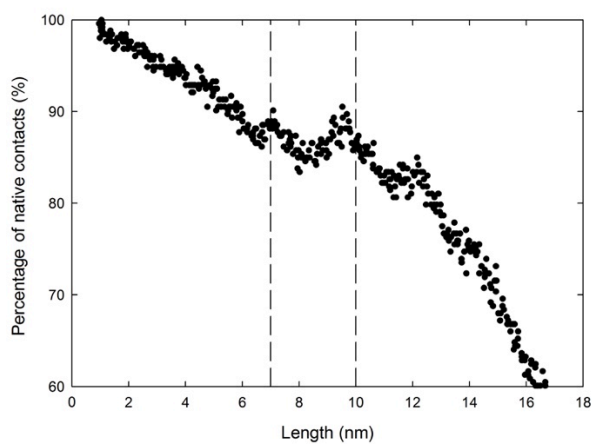

**Figure 5.** Percentage of native contacts as a function of the total length. The data correspond to the pulling of a single CC motif from its C-terminus, at a speed $3.75 \cdot 10^{-3}$ nm/ps. The borders

of the plateau agree quite well with the thresholds up to (from where) we consider the molecule to be folded (unfolded) in the geometrical picture, marked with vertical dashed lines. The analogous plot for N-pulling can be found in the supporting information.

Consistently with the above-described criteria, let us index the different subtypes of trajectories by (I, II, III, IV), attending to their degree of agreement with the theoretical prediction. G-trajectories are split into I and II subtypes: when the pulled motif is the first that unfolds, the other unit can be either still folded (type I) or partially unfolded (type II). Similarly, B-trajectories are divided into III and IV subtypes: when the non-pulled motif unfolds first, the pulled unit can be either partially unfolded (type III) or still folded (type IV). The explicit distinction between the different cases, for a simulation in which the molecules is pulled from its C-terminus, is shown in Table 1. Obviously, if the molecule is pulled from its N-terminus, the same classification of trajectories applies but with the role of the repeats C and N reversed.

**Table 1.** Definition of the different type of trajectories in a SMD C-pulling simulation. Types I and IV are the closest trajectories to a deterministic pathway, agreeing and disagreeing, respectively, with the prediction of our model.

| Type | Subtype | First repeat that unfolds | State of the other repeat | Length of the "folded" repeat |
|---|---|---|---|---|
| Good (G) | I | C-terminus | Folded | < 7 nm |
| Good (G) | II | C-terminus | Partially unfolded | > 7 nm |
| Bad (B) | III | N-terminus | Partially unfolded | > 7 nm |
| Bad (B) | IV | N-terminus | Folded | < 7 nm |

**C-pulling**

First, we pull the molecule from its C-terminus at the base velocity $v_0 = 1.4 \cdot 10^{-2}$ nm/ps. We plot the evolution of the distance between the end terminals of each repeat in this C-pulling experiment in Figure 6. The red line stands for the pulled repeat (C-terminus) whereas we plot in blue the length of the other repeat (N-terminus). It can be seen how the pulled repeat clearly unfolds first in type I. Although from different categories, types II, III and IV seem to share a common feature. In the initial part of the trajectory, it is the pulled repeat the fastest to lengthen but its unfolding comes to a standstill before being completed, and the second repeat takes advantage of this impasse to increase its extension.

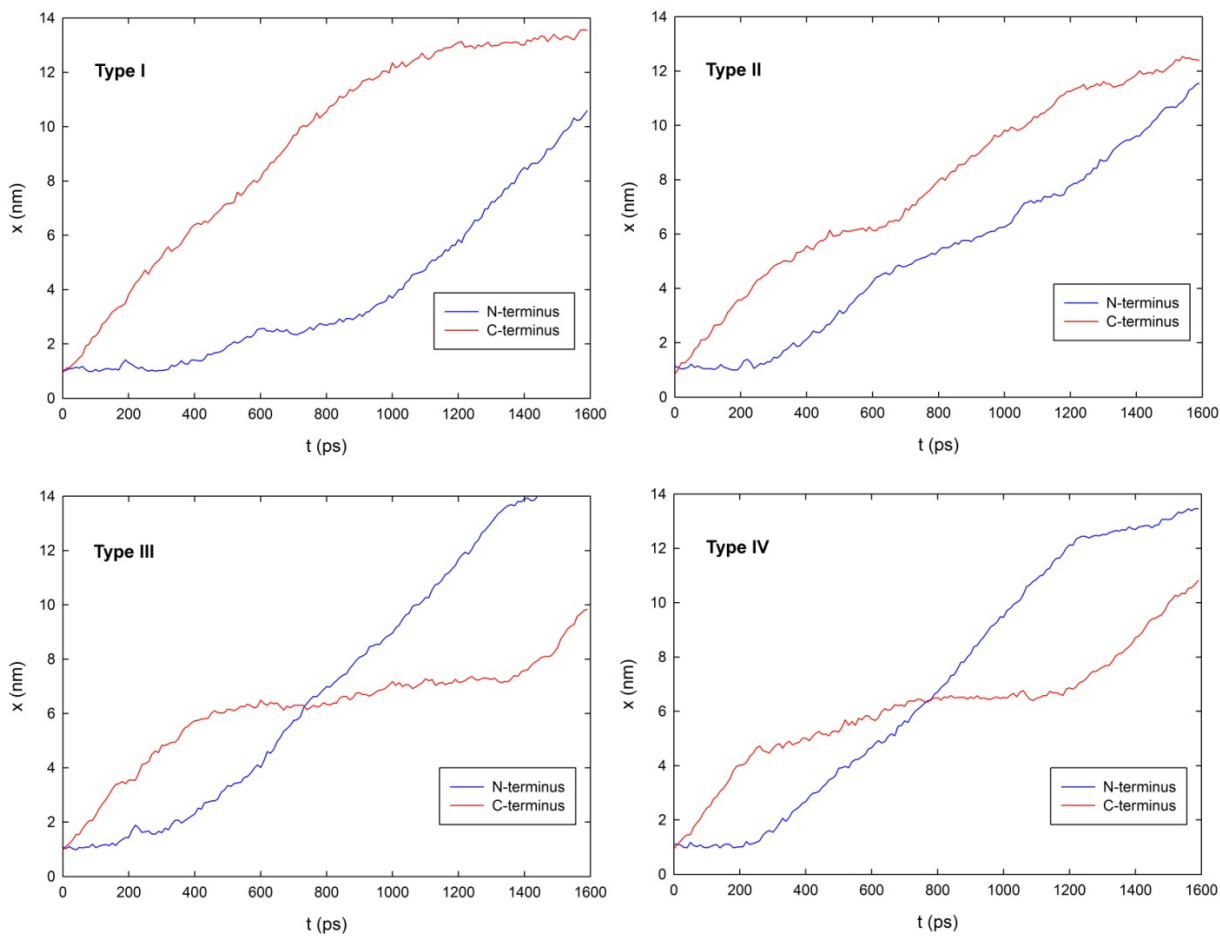

**Figure 6.** Representative plots of the different types of trajectories for SMD C-pulling simulations of the two CC construct. Each panel corresponds to a given type, as labeled. The extensions of both repeats are plotted: N-repeat (blue) and C-repeat (red). Our model predicts that the pulled repeat (C-terminus) is the first that unfolds. Videos corresponding to each of the trajectory types can be found in the supporting information that accompanies this paper.

Due to thermal fluctuations, we do not expect to obtain a perfect agreement with our theory, but a preponderance of the deterministic (type I) trajectories. As discussed before, a "fast" pulling velocity is necessary to be in the "maximum hysteresis path" limit, in which our theory is expected to hold. Therefore, in addition to the base velocity $v_0 = 1.4 \cdot 10^{-2}$ nm/ps, we have carried out simulations at velocity $2v_0$. Specifically, we have done 31 trajectories for each velocity and collected their statistical information in Table 2, which is completely compatible with our theoretical expectation. There is already a preponderance of type I trajectories at the base velocity $v_0$, at which they represent almost half of the total number of trajectories, against a reduced fraction, only 16%, of type IV trajectories. Moreover, the prevalence of type I trajectories increases with the pulling velocity, as expected in our theoretical framework: at velocity $2v_0$, more than two thirds of the SMD runs are of type I and there are no type IV trajectories.

We can get additional insight by calculating the frequencies of G (sum of I+II) and B (sum of III+IV) trajectories. This simplification of the types of events shows, even more clearly, the preponderance of the deterministic path as the velocity is increased: the frequency of G-trajectories shows a clear increase from 61% at velocity $v_0$ to 87% at velocity $2v_0$ (the frequency of B-trajectories decreases from 39% to 13%). This is the reason why we have not considered even higher velocities for C-pulling.

Our results show that increasing the pulling speed effectively diminishes the relevance of thermal activation effects and makes the unfolding more "deterministic", in the sense of increasing the prevalence of type I trajectories. It should also be stressed that detailed analysis of the rest of type trajectories highlights a branching from type I due to an impasse of the length of pulled repeat. Wrapping things up, our theory seems to predict the unfolding mechanism displayed in the SMD of this CC homopolyprotein.

**Table 2.** Statistical analysis of the output of 31 runs of SMD C-pulling simulations of the two CCs construct. At the base velocity $v_0$, there is significant preponderance of type I trajectories

(45%) as compared to type IV ones (16%). Type I trajectories clearly prevail as the pulling velocity is increased to $2v_0$, at which its frequency is boosted up to 67% whereas no type IV trajectories are found. This behavior is in clear agreement with our theoretical prediction.

| Type | Occurrence at $v_0$ (%) | Occurrence at $2v_0$ (%) |
|:---:|:---:|:---:|
| I | 45 | 68 |
| II | 16 | 19 |
| Good=I+II | 61 | 87 |
| III | 23 | 13 |
| IV | 16 | 0.0 |
| Bad=III+IV | 39 | 13 |

**N-pulling**

Of course, our one-dimensional theory is completely left right symmetric, since the free energy only depends on the extensions. Therefore, if we perform the same kind of SMD simulations but pulling from the N-terminus, we expect similar results (within statistical errors for the limited number of trajectories) and thus the unfolding to start, preponderantly, from the unit closer to the N-terminus. Nevertheless, we show below that the situation is more complex, which we understand as a signature of anisotropy in the molecule.

Table 3 presents the statistics of the different types of trajectories for SMD simulations in which we pull from the N-terminus. In terms of the same base velocity $v_0 = 1.4 \cdot 10^{-2}$ nm/ps, we have conducted numerical experiments at velocities $v_0$, $2v_0$ and $5v_0$. Specifically, the statistics have been obtained again from 31 runs for each pulling velocity. The situation is much more complex than for C-pulling, since for both $v_0$ and $2v_0$ there is no clear preponderance of the deterministic pathway. Although the fraction of G-trajectories increases from less than half to more than two thirds as the pulling velocity is increased from $v_0$ to $2v_0$, which seems to indicate that the pathway is becoming more deterministic, the frequency of type I trajectories decreases in favor of those of type II. This complex behavior makes it necessary to consider a higher velocity, so as to ascertain the tendency of the pathway as the velocity is increased. Specifically, we have incorporated simulations with pulling velocity $5v_0$. For this velocity, almost every trajectory (97%) is type I, again in agreement with our theoretical framework.

**Table 3.** Statistical analysis of the output of 31 runs of SMD N-pulling simulations of the two CCs construct. In addition to the pulling velocities $v_0$ and $2v_0$, already analyzed for C-pulling, we have considered a faster velocity $5v_0$.

| Type | Occurrence at $v_0$ (%) | Occurrence at $2v_0$ (%) | Occurrence at $5v_0$ (%) |
|---|---|---|---|
| I | 29 | 13 | 97 |
| II | 13 | 55 | 3 |
| Good=I+II | 42 | 68 | 100 |
| III | 23 | 29 | 0 |
| IV | 35 | 3 | 0 |
| Bad=III+IV | 58 | 32 | 0 |

Anisotropy of biological systems has been extensively studied[37]. Indeed, in the work of Gao et al.[38], a coiled-coil system very similar to ours presents different unfolding kinetics depending on the direction of the pulling: N- or C-pulling. Therein, the observed N-pulling transition rates between folded and unfolded states were much higher than the C-pulling rates. This property is compatible with our observations: if transition rates are higher in N-pulling, thermally activated jumps from the folded to the unfolded state could be relevant for the slower pulling velocities, although they were not for C-pulling. In this situation, it is reasonable to expect the range of velocities in which the "deterministic" or "maximum hysteresis" prevails to depend on the end from which the molecule is pulled and, specifically, be higher for N-pulling. Anyhow, the deterministic pathway becomes preponderant when the pulling velocity is high enough, as confirmed by our N-pulling simulations with the fastest velocity $5v_0$.

CONCLUSIONS

We have put forward a simple theoretical model that predicts the unfolding pathway of homopolyproteins that are pulled from one of their endpoints. Our predictions have been verified by means of all-atom steered molecular dynamics simulations in a very simple system composed of two repeats of an antiparallel CC motif.

The aforementioned model gives a mathematical foundation to the physical intuition about how a pulling force acts on a macromolecule that is being stretched. According to our theory and simulation results for high enough pulling speeds, the pulled repeat in a homopolyprotein is the fastest to increase its length and thus the first to reach the unfolded state. This new insight improves our understanding about the dependence on the pulling direction and speed of the unfolding pathway of polyproteins.

The chosen molecule and range of velocities play a central role for testing the theory. For instance, for our first considered velocity, $v_0 = 1.4 \cdot 10^{-2}$ nm/ps, our simulation results strongly depend on the terminus from which the molecule is pulled. In C-pulling simulations, there is already a preponderance of deterministic (type I) trajectories, whereas none of the trajectory types clearly prevail in N-pulling simulations. As the velocity is increased to $2v_0$, the fraction of trajectories agreeing with the deterministic path is overwhelming for C-pulling, but this is not the case for N-pulling. An even faster pulling speed, specifically $5v_0$, has had to be considered for N-pulling in order to obtain a neat predominance of the deterministic pathway.

On the one hand, the increase of the preponderance of the deterministic path with faster pulling speeds is consistent with our theoretical approach. The deterministic path is found when thermal fluctuations are effectively suppressed by the fast pulling speed, and the system does not have enough time to surpass the free energy barrier separating the folded and unfolded states. On the other hand, both the frequencies for the various trajectory types and the velocities needed to reach the deterministic path differ in C- and N-pulling. This complex behavior can be understood as stemming from the molecule being anisotropic, analogously to the reported investigation of a similar structure[38].

Quite recently, experiments and simulation have shown that the unfolding of proteins when translocated through nanopores is also a multistep process[39–41], similarly to the "classic" case of MBP studied by Bertz and Rief[18]. Specifically, an unfolding pathway that strongly depends on the pulling direction has been found in the translocation of thierodoxin through an α-haemolysing pore[39]. Since the pulling in these experiments is basically one-dimensional, it is tempting to claim that our results may be extended to this field. However, the geometrical aspects of the pulling process are quite different in AFM and co-translocational unfolding, see for instance Figure 1 of Ref.[41]. In order to apply our theory to co-translocational unfolding, we should have to incorporate the interaction between the protein and the pore, which introduces further geometrical considerations that have been disregarded in our theory. Anyhow, this seems a promising field to build a more elaborate theoretical approach.

There are several possible avenues of future research. First, our results are limited by the range of pulling speeds employed in molecular dynamics simulations, which are several orders of magnitude above those typical of pulling experiments. In this sense, an interesting prospect is finding modular proteins for which the "maximum hysteresis" or "deterministic" path lies within the range of experimental pulling speeds. Second, so far, experimental control over the pulling direction and the unfolding order of the domains is not completely robust. However, recent techniques[42], which use specific interactions for attaching the pulled macromolecule, may be used to improve the control on the pulling direction; thus opening the door to experimentally testing our theory. Finally, a more general theory predicts the unfolding order of more complex polyproteins comprising repeats with different stability properties[23]. Therein, a set of critical velocities emerged: when the pulling speed crosses these critical velocities, the first repeat that unfold changes. Then, another perspective is testing this theoretical picture in some simple macromolecules, like a modular protein composed by identical repeats except for a mutated one.

ASSOCIATED CONTENT

**Supporting Information**.

This manuscript is accompanied by several files as supporting information:

1) A file containing more details on the molecular dynamics simulation procedure, the check of the independence of the initial conformations chosen for the pulling stage, and some additional comments on the unfolding thresholds and the found anisotropic features (Supporting-info.pdf)

2) Four files containing movies of the unfolding process (C-pulling), corresponding to the four trajectory subtypes (I, II, III and IV). (typeX.m4v, where X has to be substituted by the corresponding trajectory subtype, from I to IV.)

# AUTHOR INFORMATION

**Corresponding author**

*E-mail: prados@us.es Phone: (+34) 954 55 95 14 Fax: (+34) 954 55 09 44

**Author contributions**

A. P. and P. E. M. designed the research project, A. P. and C. A. P. developed the theoretical model with feedback from P. E. M. and Z. N. S., C. A. P. performed and analyzed the computer simulations with help from Z. N. S. and all the authors equally contributed to the writing of the manuscript.

**Notes**

The authors declare no competing financial interest.

# ACKNOWLEDGMENTS

A. P. and C. A. P. acknowledge the support of the Spanish Ministerio de Economía y Competitividad through Grant FIS2014-53808-P. C. A. P. also acknowledges the support from the FPU Fellowship Programme of the Spanish Ministerio de Educación, Cultura y Deporte through Grant FPU14/00241 as well as from its Fellowship associated for research stays EST15/00051. P. E. M. and Z. N. S. acknowledge the support of National Science Foundation through grant MCB-1517245.

# SUPPORTING INFORMATION FOR "RELEVANCE OF THE SPEED AND DIRECTION OF PULLING IN SIMPLE MODULAR PROTEINS"

*Carlos A. Plata[1], Zackary N. Scholl[2], Piotr E. Marszalek[3], and Antonio Prados[1]\**

[1] Física Teórica, Universidad de Sevilla, Apdo. de Correos 1065, Sevilla 41080, Spain.

[2] Department of Physics, University of Alberta, Edmonton, Alberta, Canada.

[3] Department of Mechanical Engineering and Materials Science, Duke University, Durham, North Carolina.

In this supporting information, we include some results that are not strictly necessary for the understanding of the conclusions of the paper, but which give complementary insight that may be useful for some readers. In the first section, Simulation Details, we provide a more detailed account of all the parameters and choices in the steered molecular dynamics simulations, in order to make it possible to completely reproduce our results. The second section, Independence of the Initial Conformations in the Pulling Stage, is devoted to the analysis of the frequency of couples of unfolding pathways in consecutive simulations, in order to show that our choice of initial conformations for the pulling stage do not introduce any correlations between the unfolding events. We show that the frequencies for these couples of events are utterly compatible with the frequencies of individual unfolding events, assuming that they are independent, within our numerical accuracy. The third section, Anisotropy and Unfolding Criterion, discusses if the anisotropy found in the unfolding of the molecule has some impact on the criterion that we have used for defining folded/partially unfolded/completely unfolded states. We show that this is not the case, i.e. the unfolding thresholds used are equally suitable for C- and N-pulling. Finally, in the fourth section Videos of the Unfolding Process, we list some videos that accompany this supporting information.

SIMULATION DETAILS

In order to keep this section self-contained, we give here many details that are already in the main text. Our aim is to facilitate the reader potentially interested in reproducing or building on our results to have all the ingredients to do so, without having to look for them in different parts of the manuscript.

The homopolyprotein that we have constructed for being pulled is a simple concatenation of two coiled-coil (CC) motives. The structure of an antiparallel coiled-coil motif (CC) is extracted from the archeal box C/D sRNP core protein (Protein Data Bank entry 1NT2), which contains 67 residues and their N-terminus and C-terminus residues are arginine and isoleucine, respectively[1]. We employ this motif to build a modular protein with two units, by concatenating two identical CC motives. This is done by introducing a linker connecting them, with the linker being simply composed of two consecutive pairs of alternate residues of glycine and serine. Figure 1 of this supporting information shows the initial conformation of our model molecule[*]. The $x$-axis is chosen to be aligned with the end-to-end vector pointing from the N-terminus to the C-terminus, and the $z$-axis as parallel as possible to the common axial direction of the two motives.

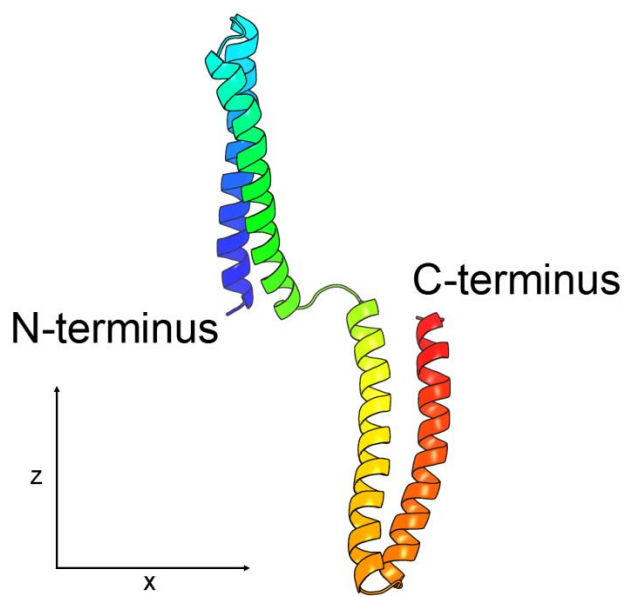

**Figure 1.** Initial conformation of the homopolyprotein composed of two CCs employed in the steered molecular dynamics simulations.

---

[*] This figure is identical to Figure 1 of the main text and is included here to keep this section self-contained, as stated in its beginning paragraph.

Some preparation is needed before the molecular dynamics simulations are implemented. Starting from the conformation in Figure 1, we use VMD Automatic PSF Builder[2] to add hydrogen atoms and afterwards create a water box of size 300 Å × 70 × Å × 120 Å, in the $x$-, $y$-, and $z$-axes, respectively. The aspect ratio of the box has been chosen to be big enough to contain the molecule, even when completely unfolded; this is why the size in the direction of the $x$-axis is the longest. Besides, we replace some water molecules by NaCl, specifically until its concentration reaches 150 mM/L and the charge is neutralized.

After the procedure described in the previous paragraph has been completed, we start the molecular dynamics simulations. To do so, we have employed NAMD2 2.10[3]. Firstly, we carry out the *equilibration stage* at 310 K. We have built a long trajectory, with two well-differentiated parts, in this stage. In the first part, whose duration is 1.5 ns, we have left the system reach equilibrium at the considered temperature. In the second part, we have picked and stored the conformations of the molecule at several times $t_k$, with $t_{k+1} - t_k \geq 0.1$ ns, $k = 1, \dots, N_T$. We denote these conformations by $\Gamma_k$.

Secondly, these different equilibrium conformations $\Gamma_k$ have been used as the initial conditions of the molecule in the $N_T$ trajectories carried out in the subsequent "pulling stage". In our simulations, the molecule is pulled by introducing an additional "ghost" atom, the position of which is perfectly controlled: it moves at constant speed $v_p$ along the $x$-axis. This ghost atom is harmonically linked with the Cα atom of the corresponding terminus (C- or N-) from which we want to pull. The force constant of this harmonic interaction is 4860 pN/nm.

The pulling stage has to be long enough so as to have the molecule completely unraveled at the final time. Therefore, the duration of the pulling stage $\Delta t_p$ is fixed by the size of the considered molecule in its axial ($z$-axis) direction. In our case, this length is 4.82 nm, which corresponds to the separation between the two C$\alpha$ atoms of the residues that are most distant in the axial direction (specifically, ARG and ILE). Therefore, we have chosen $\Delta t_p$ such that the product $v_p \Delta t_p >$20 nm, since the length of a completely unfolded motif is approximately 10 nm. Specifically, we have chosen $\Delta t_p = \Delta t_0 =$1.6 ns for our "base" velocity $v_0$, which leads to $v_0 \Delta t_0 =$22.4 nm and, in general, $\Delta t_p = \Delta t_0\, v_0/v_p$ for a pulling velocity $v_p$.

INDEPENDENCE OF THE INITIAL CONFORMATIONS IN THE PULLING STAGE

The intention of the above described procedure to obtain the different initial conformations $\Gamma_k$ is to generate a sample of "independent" initial conditions for the pulling stage. What we mean by independent is that the unfolding pathway from these conformations should not be correlated. But, is this so?

In order to check the independence of the initial conformations $\Gamma_k$ in the sense explained in the above paragraph, we have analyzed possible correlations between the observed pathways in consecutive runs $k$ and $k + 1$. In particular, we have focused our attention on the four possible pairs of "good" (G) and "bad" (B) trajectories (GG, GB, BG, BB) for each case we have considered, i.e. for given pulled terminus and pulling speed. We recall that G-trajectories correspond to those in which the pulled unit unfolds first, consistently with our theory.

The $k$-th pulling trajectory, for given pulled terminus and pulling speed, starts from conformation $\Gamma_k$. Let us introduce a stochastic variable $\xi_k$ to identify the type to which the $k$-th trajectory belongs. Thus, $\xi_k$ can have two values that are denoted by $\alpha$, i.e. $\alpha$ is equal to either G or B. In consecutive runs, $k$ and $k + 1$, the possible values of the pair $(\xi_k, \xi_{k+1})$ are denoted by $\alpha\beta$, i.e. the four possibilities (GG, GB, BG, BB). We denote the number of pairs of consecutive trajectories with a certain outcome $\alpha\beta$ by $n_{\alpha\beta}$, which is

$$n_{\alpha\beta} = \sum_{k=1}^{N_p} \delta_{\xi_k,\alpha} \, \delta_{\xi_{k+1},\beta}, \qquad (1)$$

where $N_p = N_T - 1$ is the number of consecutive pairs and $\delta_{ij}$ is Kronecker's delta.

The probability that any trajectory corresponds to a given type (G or B) can be formally written as

$$\langle \delta_{\xi_k,\alpha} \rangle = p_\alpha. \qquad (2)$$

We are denoting the corresponding probabilities by $p_G$ and $p_B$, respectively, and $p_G + p_B = 1$ because these events are mutually exclusive, $\delta_{\xi_k,G} = 1 - \delta_{\xi_k,B}$. Now, assuming the outcomes of consecutive trajectories to be independent, we can ask ourselves the following questions in an ensemble of simulations comprising $N_T$ trajectories (corresponding to a given pulled terminus and pulling speed).

(i) What is the expected number of each pair type $\langle n_{\alpha\beta} \rangle$?

(ii) What are their corresponding standard deviations $\sigma_{\alpha\beta}$?

The mean value is directly obtained by taking into account eq. (2), for all $k$, and the assumed statistical independence of the variables $\xi_k$ and $\xi_{k+1}$. Therefore, one gets straightforwardly that

$$\langle n_{\alpha\beta} \rangle = N_p p_\alpha p_\beta. \qquad (3)$$

The derivation of the expression for the fluctuations is lengthier. We start by writing

$$n_{\alpha\beta}^2 = \sum_{k=1}^{N_p} \sum_{l=1}^{N_p} \delta_{\xi_k,\alpha}\, \delta_{\xi_{k+1},\beta}\, \delta_{\xi_l,\alpha}\, \delta_{\xi_{l+1},\beta}, \qquad (4)$$

and we split the above expression into four contributions, corresponding to the cases $l = k$, $l = k \pm 1$, and all the other values of $l$. Thus, we can write

$$n_{\alpha\beta}^2 = \sum_{k=1}^{N_p} \delta_{\xi_k,\alpha}\, \delta_{\xi_{k+1},\beta} + \delta_{\alpha\beta} \sum_{k=1}^{N_p-1} \delta_{\xi_k,\alpha}\, \delta_{\xi_{k+1},\alpha} \delta_{\xi_{k+2},\beta}$$
$$+ \delta_{\alpha\beta} \sum_{k=2}^{N_p} \delta_{\xi_{k-1},\alpha}\, \delta_{\xi_k,\alpha} \delta_{\xi_{k+1},\beta} + \sum_{k=1}^{N_p} \sum_{l \neq k, k\pm 1}^{N_p} \delta_{\xi_k,\alpha}\, \delta_{\xi_{k+1},\beta}\, \delta_{\xi_l,\alpha}\, \delta_{\xi_{l+1},\beta}. \qquad (5)$$

Now we compute the average. We have four contributions on the rhs of the equation above: in the first sum, there are $N_p$ terms, each of them with average $p_\alpha p_\beta$; in both the second and third sums, there are $(N_p - 1)$ terms, each of them with average $p_\alpha^2 p_\beta$; and in the fourth sum we have the remaining $N_p^2 - (3N_p - 2)$ terms, each of them with average $p_\alpha^2 p_\beta^2$. Once more, we have assumed the independence of the variables $\xi_k, \xi_l$ for $l \neq k$. Thus, we get

$$\langle n_{\alpha\beta}^2 \rangle = N_p p_\alpha p_\beta + 2\delta_{\alpha\beta}(N_p - 1) p_\alpha^2 p_\beta + [N_p^2 - (3N_p - 2)] p_\alpha^2 p_\beta^2. \qquad (6)$$

The variance is given by

$$\sigma_{\alpha\beta}^2 \equiv \langle n_{\alpha\beta}^2 \rangle - \langle n_{\alpha\beta} \rangle^2 = N_p p_\alpha p_\beta (1 - 3 p_\alpha p_\beta) + 2 p_\alpha^2 p_\beta^2 + \delta_{\alpha\beta} 2(N_p - 1) p_\alpha^2 p_\beta. \qquad (7)$$

For the sake of clarity, we list below the average values and the variances for the three pairs of outcomes leading to different values; note that $\langle n_{GB} \rangle = \langle n_{BG} \rangle$ and also $\sigma_{GB} = \sigma_{BG}$. Specifically,

$$\langle n_{GG} \rangle = N_p p_G^2, \qquad \sigma_{GG}^2 = p_G^2 (1 - p_G)[N_p + p_G (3N_P - 2)], \qquad (8a)$$

$$\langle n_{BB}\rangle = N_p p_B^2, \qquad \sigma_{BB}^2 = p_B^2(1-p_B)[N_p + p_B(3N_p - 2)], \tag{8b}$$

$$\langle n_{GB}\rangle = \langle n_{BG}\rangle = N_p p_G p_B, \qquad \sigma_{GB} = \sigma_{BG} = N_p p_G p_B(1 - 3p_G p_B) + 2p_G^2 p_B^2. \tag{8c}$$

As expected, the variances would vanish if the process were purely deterministic and $p_G = 1$ or, equivalently, $p_B = 0$, since all the pairs would correspond to the $GG$ case.

From an empirical point of view, we can identify $p_G$ and $p_B$ with the frequencies of G- and B-trajectories for the considered experiment (given pulled terminus/pulling speed). After doing that, we can count the actual number of pairs $n_{\alpha\beta}$ for each pair type in the ensemble of $N_T$ trajectories, and check if it lies within the theoretical expectation, for these empirical values of $p_\alpha$. What we show below is that for all our simulations we have that $|n_{\alpha\beta} - \langle n_{\alpha\beta}\rangle| \leq \sigma_{\alpha\beta}$, i.e. the agreement, considering the initial conformations as independent, is really good.

In order to give all the results in only one table, we give the empirical frequencies for each ensemble of trajectories, corresponding to a given pulled terminus and pulling speed. For example, C-$v_0$ means that the data in the corresponding column is for C-pulling at velocity $v_0$. For each column, the rows correspond to the frequencies for different events: (i) in the G and B rows, we give the empirical frequencies for good and bad trajectories, taken from Tables 2 and 3 in the main text; (ii) in the subsequent $\alpha\beta$ rows, we list the empirical frequencies for two consecutive trajectories of type $\alpha\beta$; (iii) in the $\alpha\beta$-th row, we give the theoretical prediction for that frequency, calculated as $\langle n_{\alpha\beta}\rangle/N_p$ from eqs. (8); (iv) in the $\alpha\beta$-err, we provide the theoretical prediction for the standard deviation of that frequency, calculated as $\sigma_{\alpha\beta}/N_p$ from eqs (8). For the theoretical values of $\langle n_{\alpha\beta}\rangle/N_p$ and $\sigma_{\alpha\beta}/N_p$, we take $p_G$ and $p_B$ equal to the empirical frequencies of G- and B-trajectories, respectively, as already said above.

**Table 1.** Frequencies (empirical and theoretical) and expected statistical errors for individual events and pair of consecutive events. The values for the frequencies of G- and B-trajectory types are taken from Tables 2 and 3 of the main text.

| Frequencies | C-$v_0$ | C-$2v_0$ | N-$v_0$ | N-$2v_0$ | N-$5v_0$ |
|---|---|---|---|---|---|
| G | 0.61 | 0.87 | 0.42 | 0.68 | 1.00 |
| B | 0.39 | 0.13 | 0.58 | 0.32 | 0.00 |
| GG | 0.37 | 0.73 | 0.13 | 0.50 | 1.00 |
| GG-th | 0.38 | 0.76 | 0.18 | 0.46 | 1.00 |
| GG-err | 0.12 | 0.10 | 0.09 | 0.12 | 0.00 |
| GB | 0.23 | 0.13 | 0.27 | 0.17 | 0.00 |
| GB-th | 0.24 | 0.11 | 0.24 | 0.22 | 0.00 |
| GB-err | 0.05 | 0.05 | 0.05 | 0.05 | 0.00 |
| BG | 0.23 | 0.13 | 0.27 | 0.17 | 0.00 |
| BG-th | 0.24 | 0.11 | 0.24 | 0.22 | 0.00 |
| BG-err | 0.05 | 0.05 | 0.05 | 0.05 | 0.00 |
| BB | 0.17 | 0.00 | 0.33 | 0.17 | 0.00 |
| BB-th | 0.15 | 0.02 | 0.34 | 0.10 | 0.00 |
| BB-err | 0.08 | 0.02 | 0.11 | 0.07 | 0.00 |

In our study on the independence of the initial conformation, we have restricted ourselves to the coarse-grained classification into G- and B-trajectory types, instead of the more precise subtypes I to IV. The reason behind this choice is the limited number of trajectories considered, $N_T = 31$, for each pulled terminus and pulling speed, which are not enough to analyze the 16 possibilities arising for the outcome in two consecutive trajectories. Also, we would like to stress that, if the pathways found from conformations $\Gamma_k$ and $\Gamma_{k+1}$ are not correlated, nor will the pathways from $\Gamma_k$ and $\Gamma_{k+n}$, with $n > 1$, be. This is clear from a physical point of view, what we have shown is that the minimum time between consecutive conformations in the equilibration stage is longer than some characteristic time, beyond which correlations between the molecule conformations are lost.

ANISOTROPY AND UNFOLDING CRITERION

In the main text, we have shown that the molecule is anisotropic, in the sense that the unfolding pathway that we find for a given pulling speed depends on the terminus from which the molecule is pulled. Specifically, the frequencies of either the different trajectory types (G or B) or subtypes (I, II, III or IV) for a given pulling speed are different for C-pulling than those for N-pulling.

Our unfolding criterions/thresholds for classifying the molecule as folded/partially unfolded/completely unfolded are based on a geometrical picture, with further support from our analysis of the fraction of native contacts as a function of the length (see figure 5 of the main text). However, this last analysis was done by using a C-pulling simulation and, given the revealed anisotropic features of the molecules, one may wonder whether the unfolding criterions/thresholds depend on the pulling direction.

In order to answer the above issue, we have carried out a N-pulling simulation with the same pulling speed used in figure 5 of the main text. We present the results of such a simulation in figure 1, in which the percentage of native contacts is plotted against the end-to-end length of the molecule. This N-pulling curve is superposed with that from the C-pulling simulation and it is observed that, aside from the latter being noisier than the former, the two curves give the same estimates for the unfolding thresholds.

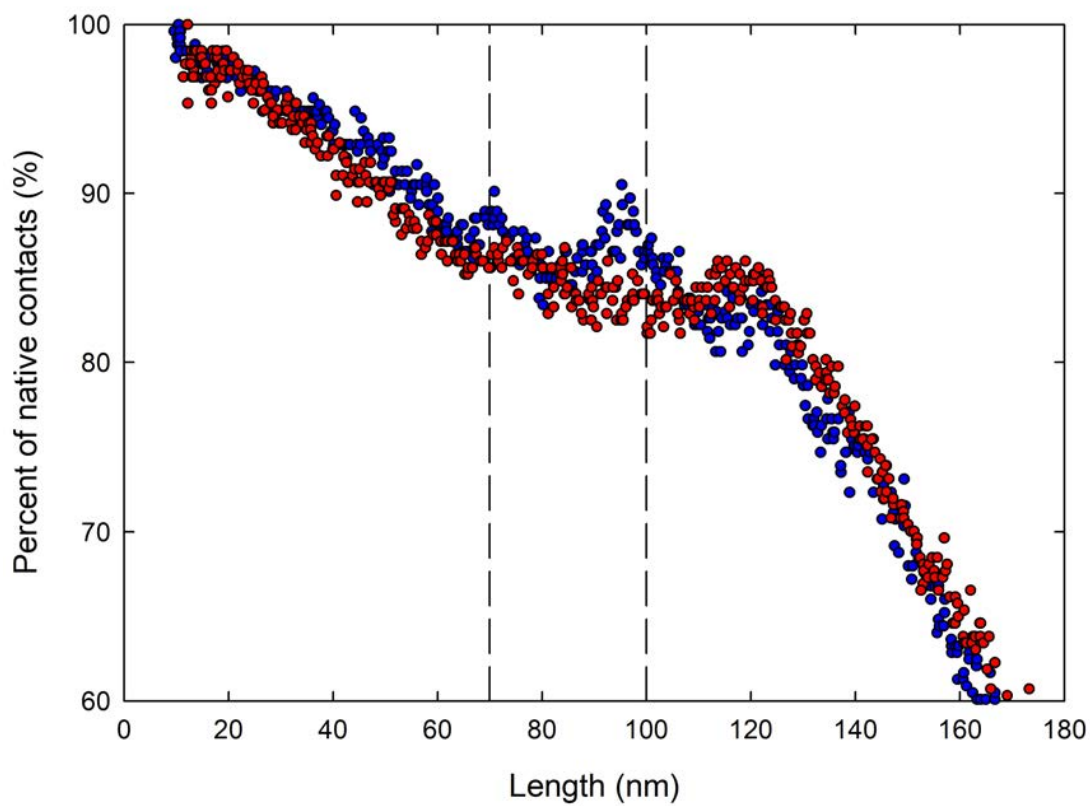

**Figure 1.** Percentage of native contacts as a function of the system length. Both the data from C-pulling (blue) and N-pulling (red) are plotted. The same pulling speed has been used for the two pulling directions, namely $3.75\times10^{-3}$ nm/ps.

VIDEOS OF THE UNFOLDING PROCESS

Further insight into the unfolding process can be gained by looking at four videos we have prepared and that accompany this supporting information file. All of them correspond to C-pulling at the "base" velocity $v_0 = 1.4 \cdot 10^{-2}$ nm/ps and show the typical behavior of the molecule in each of the four subtypes of trajectories. They are named accordingly as typeI.m4v, type2.m4v, typeIII.m4v and typeIV.m4v.